\documentclass[reprint,aps,prl,nofootinbib,superscriptaddress,nobalancelastpage]{revtex4-2}
\usepackage{amsmath}
\usepackage{amsfonts}
\usepackage{amssymb}
\usepackage{amsthm}
\usepackage{mathtools}
\usepackage{graphicx}
\usepackage[dvipsnames]{xcolor}
\usepackage[colorlinks=True,citecolor=myBlue,linkcolor=myRed,urlcolor=myBlue,hypertexnames=false]{hyperref}
\usepackage{upgreek}
\usepackage{hypcap}
\usepackage{yfonts}
\usepackage{bm}
\usepackage[normalem]{ulem}
\usepackage{dsfont}
\usepackage{braket}
\usepackage{float}
\usepackage{marginnote}
\usepackage{tikz}
\usepackage{pgfplots}
\pgfplotsset{compat=1.14}
\usetikzlibrary{shapes, arrows.meta,decorations.pathreplacing}

\newcommand{\nh}{N_{\cal H}}
\newcommand{\abs}[1]{|#1|}

\newcommand\eq[1]{\begin{align}#1\end{align}}

\newcommand{\pd}{\phantom\dagger}
\definecolor{myBlue}{RGB}{31,119,180}
\definecolor{myOrange}{RGB}{255,127,14}
\definecolor{myGreen}{RGB}{44,160,44}
\definecolor{myRed}{RGB}{214,39,40}
\definecolor{myPurple}{RGB}{148,103,189}

\newcommand\mytitle{
Statistics of Fock-space correlation lengths in the many-body localised phase}

\begin{document}

\title{\mytitle}

\author{Bibek Saha}
\email{bibek.saha@icts.res.in}
\affiliation{International Centre for Theoretical Sciences, Tata Institute of Fundamental Research, Bengaluru 560089, India}

\author{David E. Logan}
\email{david.logan@chem.ox.ac.uk}
\affiliation{
University of Oxford, Physical and Theoretical Chemistry, South Parks Road, Oxford OX1 3QZ, UK
}

\author{Sthitadhi Roy}
\email{sthitadhi.roy@icts.res.in}
\affiliation{International Centre for Theoretical Sciences, Tata Institute of Fundamental Research, Bengaluru 560089, India}

\begin{abstract}
The presence of long-ranged, many-body resonances in the many-body localised phase of disordered quantum systems is manifest in strong inhomogeneities of eigenstate amplitudes on the Fock space. We develop a quantitative, microscopic theory for this, centred on two-point Fock-space correlation functions. Exact diagonalisation of the disordered tilted-field Ising chain uncovers a stark contrast: typical correlations decay perfectly exponentially, whereas average correlations exhibit subexponential decay with systematic system-size dependence driven by rare 
resonant nodes. We show that this behaviour stems from an underlying distribution of Fock-space correlation lengths with a variance that scales inversely with system size. Crucially, the interplay of this distribution with the high-dimensional geometry of the hypercubic Fock-space graph, leads to the qualitative difference between the typical and average behaviours. The theory also allows us to derive the singularity spectrum of the many-body localised eigenstates, and to establish theoretically the anomalous symmetry violation of quantum multifractality as an inherent consequence of the Fock-space geometry.
\end{abstract}

\maketitle


\paragraph*{Introduction:}
The many-body localised (MBL) phase in strongly disordered, interacting quantum many-body systems presents a robust paradigm for the breakdown of thermalisation in isolated systems~\cite{basko2006metal,gornyi2005interacting,oganesyan2007localisation,nandkishore2015many,abanin2019colloquium,sierant2025many}.
A useful perspective to understand the MBL phase is that it is smoothly connected to the non-interacting Anderson localised phase via a finite-depth local unitary operator, leading to the emergence of an extensive number of quasilocal integrals of motion~\cite{bauer2013area, serbyn2013local, huse2014phenomenology}.
Indeed, this led to an understanding of several features of the MBL phase, such as the absence of transport and the logarithmic growth of entanglement entropy~\cite{bardarson2012unbounded,serbyn2013universal,serbyn2014quantum}.
However, more recently it has become clear that this picture is not entirely complete, as it does not capture the physics of long-ranged many-body resonances, which are an essential feature of the MBL phase~\cite{garratt2021resonances,morningstar2021avalanches, garratt2022resonant, crowley2022constructive, biroli2024largedeviation,colbois2024, laflorencie2025cat}.

Concurrently, an insightful alternative viewpoint that has emerged for the MBL problem is to study the interacting system as an effective single-particle problem on the disordered, correlated, high-dimensional Fock-space graph (see Refs.~\cite{tikhonov2021anderson,roy2024fock} for reviews and references therein).
By focussing on the anatomy of eigenstates on Fock space~\cite{tikhonov2021eigenstate,roy2021fockspace,detomasi2020rare}, this approach has led to fundamental insights, such as the multifractality of MBL eigenstates~\cite{deluca2013ergodicity,mace2019multifractal, roy2025spectral}, the emergence of Fock-space length scales which relate to real-space dynamical observables~\cite{detomasi2020rare,roy2021fockspace,creed2023probability}, and the key role played by disorder correlations in stabilising the MBL phase~\cite{logan2019many,roy2020fock}.
 
\begin{figure}
\includegraphics[width=\linewidth]{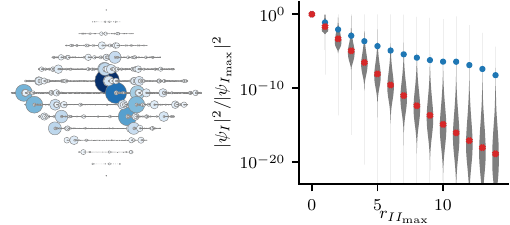}
\caption{Inhomogeneities of eigenstate amplitudes on the Fock-space graph. The left panel shows an exemplary eigenstate on the hypercubic Fock-space graph of the tilted-field Ising chain. The graph is organised such that the apical node is the localisation centre, 
$I_{\rm max}$, of the state, and the rows are ordered by their Hamming distance, 
$r_{II_{\rm max}}$, from it.  The sizes (and colour intensities) of the nodes reflect the wavefunction amplitudes relative to the average amplitude on that row.
The right panel shows the average (blue) and typical (red) intensities relative to the maximum, as a function of $r_{II_{\rm max}}^{\pd}$. The violin plots \cite{hintze1998violin}  (grey shaded regions) denote the full distribution of the intensities.}
\label{fig:fs}
\end{figure}

In this Letter, we consider the broad question of how rare, long-ranged resonances are imprinted onto the anatomy of MBL eigenstates on the Fock-space graph.
Specifically, we address the question of strong inhomogeneities of MBL eigenstates on
the Fock-space graph~\cite{roy2021fockspace}, as illustrated in Fig.~\ref{fig:fs}
(see also Fig.\ 13 of~\cite{roy2021fockspace} for further characterisation of them).
 It is important to understand the origins of these inhomogeneities, and develop a quantitative theory for them, as a stepping stone towards a microscopic understanding of the resonances. 
In particular, as shown in Fig.~\ref{fig:fs}, while the amplitudes generally decay away from the localisation centre, they are distributed rather broadly; their typical and average values 
being governed respectively by the bulk and the tails of the distribution.

In the remainder of this work, we develop a quantitative theory for these inhomogeneities, rooted in a distribution of correlation lengths on the Fock space. 
We show that this distribution is well-behaved, with a variance that decays inversely with system size. Crucially, however, the interplay of this distribution with the high-dimensional
connectivity of the Fock-space graph, underlies the strong qualitative difference between the average and typical behaviours on the Fock space.

\paragraph*{Model:}
As a concrete model, we consider the disordered, tilted-field Ising (TFI) chain, described by the Hamiltonian~\cite{imbrie2016many,Imbrie2016PRL,abanin2021distinguishing}
\eq{\label{eq:hamiltonian}
{\cal H} =\sum_{i=1}^{L-1} J_i^{\pd} \sigma_i^z \sigma_{i+1}^z +\sum_{i=1}^L [h_i^{\pd} \sigma_i^z + \Gamma \sigma_i^x] \,,
}
where (taking $\Gamma=1$) the couplings $J_i$ are drawn uniformly from $[J-\Delta J, J+\Delta J]$ with $J=1$ and $\Delta J=0.2$, and the i.i.d.\ longitudinal fields $h_i\in [-W,W]$. 
For system sizes accessible to exact diagonalisation, the model hosts an MBL phase for $W\gtrsim 4$~\cite{abanin2021distinguishing,roy2021fockspace} in the middle of the spectrum.  We restrict our numerical studies to 
$W\gtrsim 6$ to remain deep within the MBL regime. Because the disorder couples to the $z$-component of the spins, the set of $\sigma^z$-product states, $\{\ket{I}\}$, provides a natural Fock-space basis. Any eigenstate of the Hamiltonian in Eq.~\eqref{eq:hamiltonian} can be decomposed as $\ket{\psi} = \sum_{I}\psi_I\ket{I}$.

Treating these basis states as nodes, the Fock-space graph for the TFI chain maps exactly 
to an $L$-dimensional hypercube, with $\nh =2^{L}$ nodes, each having a connectivity of 
$L$~\cite{roy2024fock}. The Hamming distance between two nodes $I$ and $J$ -- denoted by $r_{IJ}$ and defined as the number of real-space sites with differing spin configurations -- endows the graph with a natural distance metric; for the hypercube, this coincides with the shortest path length between nodes. The number of nodes at Hamming distance $r$ from any given node is
$N_r =\binom{L}{r}$.


\paragraph*{Fock-space correlation functions:}
Having established the Fock-space topology and notations, we now define the 
eigenstate correlation functions, which are the central quantities of this study.
The generalised average correlation function of order $q$ is defined as
\eq{
C^{(q)}_{\rm avg}(r,L) = \left\langle \frac{1}{\nh}\sum_I\frac{1}{N_r}\sum_{K:r_{IK}=r}\abs{\psi_I^{\pd}}^{2q}\abs{\psi_K^{\pd}}^{2q}\right\rangle\,,
\label{eq:Cavg-q}
}
which clearly averages over all pairs of Fock-space nodes $I$ and $K$ separated by a distance $r$
(and $\langle\cdots\rangle$ denotes disorder average). The strong inhomogeneity of 
eigenstate amplitudes~\cite{roy2021fockspace} on the Fock-space graph implies that the two-point weights $\abs{\psi_I}^2\abs{\psi_K}^2$ are similarly broadly distributed.
This motivates us to study also the typical correlation function, defined as 
\eq{
C^{(q)}_{\rm typ}(r,L) = \exp\left\langle\ln\left(\frac{1}{\nh}\sum_I\abs{\psi_I^{\pd}}^{2q}\abs{\psi_{I_r^{\pd}}}^{2q}\right)\right\rangle\,
\label{eq:Ctyp-q}
}
with $I_r$ a node chosen uniformly at random from the $N_r$ nodes at a distance $r$ from $I$.
For a given $I$, this random sampling ensures that the correlation function in Eq.~\eqref{eq:Ctyp-q} probes the typical behaviour of the correlations~\cite{garciamata2020two,garciamata2022critical,saha2026anderson}. For our analysis, it is convenient to rescale these correlation functions by their $r=0$ values. To this end, we define the normalised quantities
\eq{
c^{(q)}_{\rm avg/typ}(r,L) = C^{(q)}_{\rm avg/typ}(r,L) / C^{(q)}_{\rm avg/typ}(0,L)\,.
\label{eq:C-rescaled}
}
Further, $C_{\rm avg}^{(q)}$ is directly connected to the generalised $q$-th inverse participation ratio (IPR) ${\cal I}_q =\sum_{I}|\psi_{I}|^{2q}$ via 
\eq{
C_{\rm avg}^{(q)}(0,L) = \nh^{-1}\big\langle{\cal I}_{2q}^{\pd}\big\rangle \sim 
\nh^{-(\tau_{2q}^{\pd}+1)}\,,\label{eq:Cavg-IPR}
}
with $\{\tau_q\}$ the standard set of multifractal exponents.

\begin{figure}
    \centering
    \includegraphics[width=\linewidth]{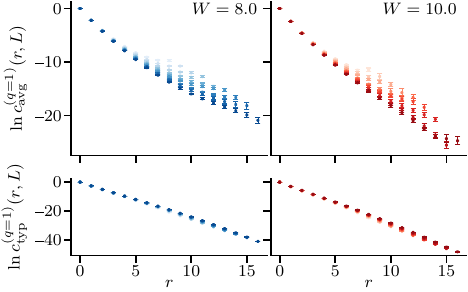}
    \caption{Average and typical correlation functions, $c_{\rm avg/typ}^{(q=1)}$ [defined in Eq.~(\ref{eq:C-rescaled})], as a function of Hamming distance $r$. While $c_{\rm avg}^{(q=1)}$ exhibits subexponential decay with non-trivial $L$-dependence, $c_{\rm typ}^{(q=1)}$ shows a pure exponential decay independent of system size $L$. Lighter to darker colours denote 
$L=8,9,\dots,16$. Data is averaged over $\sim 10^4$  disorder realisations for smaller sizes and 
$\sim 10^3$ for larger ones.
    }
    \label{fig:q=1-corrs}
\end{figure}

Fig.~\ref{fig:q=1-corrs} shows $c^{(q=1)}_{\rm avg/typ}(r,L)$ against $r$ for different $L$ at two representative values of disorder, $W=8$ and $10$, deep in the MBL phase. 
The data show that $c_{\rm avg}(r,L)$ exhibits subexponential decay with $r$ and a pronounced, systematic $L$-dependence. By contrast, $c_{\rm typ}(r,L)$ decays as a pure exponential, 
independent of $L$. 

\begin{figure}
    \centering
    \includegraphics[width=\linewidth]{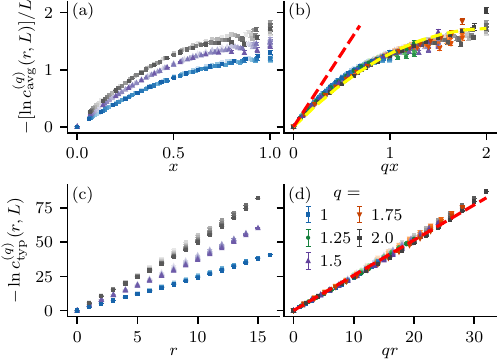}
    \caption{For $W=8$. (a) 
		Shown for $q=1,1.5, 2$, the rescaled average correlation functions for $L=8,\dots,16$ (lighter to darker colours) collapse onto common curves versus $x=r/L$. 
		(b) Data for different $q$ further collapse onto a single curve when plotted against $qx$. The yellow dashed line fits this curve to $\xi_{\rm typ}^{-1}qx - b(qx)^2$.
(c) The typical correlation functions depend purely on $r$, independent of $L$, and (d) data for different $q$ collapse onto a common exponential decay (dashed line) versus $qr$. Together, these confirm the scaling forms in Eq.~(\ref{eq:c-avg-typ-scaling}). The red dashed lines in (b) and (d) are the same.}
    \label{fig:qth-corrs-scaling}
\end{figure}

Importantly, for a given $W$, the data for different $q$ and $L$ collapse onto the scaling forms
\eq{
    \begin{split}
    c_{\rm avg}^{(q)}(r,L) &= \exp[{-L g(qx)}]\,;~~~x = r/L\,,\\
    c_{\rm typ}^{(q)}(r,L) &= \exp[{-qr/\xi_{\rm typ}}]\,,
    \end{split}
    \label{eq:c-avg-typ-scaling}
}
where $g(0)=0$ by definition in Eq.~\eqref{eq:C-rescaled}. Since we are deep in a phase, $g(z)$ must be analytic in its domain $z\in[0,1]$, implying  that it admits a series expansion 
$g(z) = \sum_{n=1}^{\infty} z^n g^{(n)}(z)/n!$. We find numerically that  $g(z)$ is sufficiently well described by the quadratic form\footnote{From Eqs.~\eqref{eq:c-avg-typ-scaling},\eqref{eq:gz}, 
$ c_{\rm avg}^{(q)}(r,L)$ can also be viewed as $\exp[-Lx/\xi_\mathrm{eff}]$ 
where $\xi_\mathrm{eff} = q^{-1} \xi_\mathrm{typ} [1 - b q \xi_\mathrm{typ} x]^{-1}$
grows with $x$, reflecting the subexponential decay of the average correlation functions.}
\eq{
g(z) \simeq \frac{z}{\xi_{\rm typ}(W)} - b(W)z^2\,;~~~b(W)>0\,.
\label{eq:gz}
}
Evidence for the universal scaling behaviour of Eq.~(\ref{eq:c-avg-typ-scaling}) is presented in Fig.~\ref{fig:qth-corrs-scaling}. 
That the slope of $g(z)$ near $z=0^+$ cleanly matches $\xi^{-1}_{\rm typ}$ can be seen from the red dashed lines in Figs.~\ref{fig:qth-corrs-scaling}(b) and (d).

These numerical results carry several key physical implications. 
First, the typical exponential decay of the $q=1$ two-point weights 
$\abs{\psi_I}^2\abs{\psi_K}^2$ (Fig.~\ref{fig:q=1-corrs})
firmly establishes a characteristic correlation length, $\xi_{\rm typ}$. 
The robustness of this length scale is corroborated by the proportional decay of the generalised weights $\abs{\psi_I}^{2q}\abs{\psi_K}^{2q}$ with a modified length scale 
$q^{-1}\xi_{\rm typ}$.
However, the qualitatively different, subexponential decay of $c_{\rm avg}^{(q)}$ indicates the existence of rare pairs of nodes where the two-point weights are anomalously large. 
While irrelevant for the typical behaviour, these rare resonant pairs completely dominate the average. This sharp contrast points to an underlying distribution of Fock-space correlation lengths, for which we now develop a theory.


\paragraph*{Distribution of Fock-space correlation lengths:} 
Since the two-point correlation typically decays exponentially but with a distribution of length scales, a natural starting point for our theory is 
\eq{
c^{(q)}_{\rm avg}(r,L) = \int d\xi\,P_{\xi}^{\pd}(\xi,L) e^{-qr/\xi}\,,
\label{eq:cavg-Pxi}
}
with $P_{\xi}(\xi,L)$ the normalised distribution of interest. 
Using the scaling form Eq.~\eqref{eq:c-avg-typ-scaling} then gives
\eq{
g(z) = -\frac{1}{L}\ln\left[\int d\xi\,P_{\xi}^{\pd}(\xi,L) e^{-Lz/\xi}\right]\,.
\label{eq:cavg-Pxiplus}
}
The coefficients $\xi_{\rm typ}^{-1}$ and $b$ in $g(z)$ [Eq.~\eqref{eq:gz}] can be directly extracted from the derivatives of $g(z)$ evaluated at $z=0$,
\eq{
\xi_{\rm typ}^{-1} = \braket{\xi^{-1}}\,; \quad b = \frac{L}{2}\left[\braket{\xi^{-2}} - \xi_{\rm typ}^{-2}\right]\,.
}
Physically, this implies that the inverse correlation length, $\xi^{-1}$, follows a distribution with mean $\xi_{\rm typ}^{-1}$ and variance ${\rm var}[\xi^{-1}] = 2b/L$. 
Because $b$ is independent of $L$ (as seen in Fig.~\ref{fig:qth-corrs-scaling}), the variance decays as $L^{-1}$. 
Consequently, in the thermodynamic limit, this distribution collapses to a Dirac $\delta$-function. 
However, as discussed next, it is precisely the non-trivial $L$-scaling of the distribution, acting in concert with the high-dimensional connectivity of the Fock-space graph, that generates the rare pairs of nodes with anomalously large correlations, ultimately  producing the subexponential decay of $c_{\rm avg}^{(q)}$.

Given the scaling form of $c_{\rm avg}^{(q)}$ in Eq.~\eqref{eq:c-avg-typ-scaling}, and since 
$\xi_{\rm typ}\sim O(1)$, a natural ansatz for the distribution is 
\eq{
P_{\xi}(\xi,L) = \frac{e^{-Lp(\xi)}}{\int d\xi\,e^{-Lp(\xi)}}\,,
}
with $p(\xi)$ intensive (and with $P_{\xi}$ manifestly normalised).
Using this  in Eq.~\eqref{eq:cavg-Pxiplus} gives
\eq{
e^{-Lg(z)} = \frac{\int d\xi\,e^{-L[p(\xi)+z/\xi]}}{\int d\xi\,e^{-Lp(\xi)}}\,.
\label{eq:F-def}
}
This large-deviation form makes it amenable to a steepest-descents analysis, which leads to
$g(z) = \min_\xi \left[p(\xi) - p(\xi_0) +z/\xi\right]$. Equivalently,
\eq{
g(z) = p(\xi_\ast(z)) - p(\xi_0)+\frac{z}{\xi_\ast(z)}\,,
}
where the saddle point $\xi_\ast(z)$ is given by 
\eq{
\partial_\xi p(\xi)\big\vert_{\xi=\xi_\ast}^{\pd}-\frac{z}{\xi^2_\ast}=0\,,
}
and with $\xi_0\equiv \xi_\ast(z=0)$.
These saddle point equations naturally imply\footnote{
Note that $g(z)=\min_\xi G(z,\xi)$, so its derivative is thus 
\eq{
g'(z) = \partial_z G(z,\xi)\vert_{\xi=\xi_\ast(z)}^{\pd} + 
\partial_\xi G(z,\xi)\vert_{\xi=\xi_\ast(z)}^{\pd}\frac{d}{dz}\xi_\ast(z)\,.
\nonumber
}
But the second term on the RHS is zero, as $\xi_\ast(z)$ is a minimum of $G(z,\xi)$. Using 
$G(z,\xi)= \tilde{p}(\xi)+z/\xi$ gives  $g'(z)=\xi_\ast^{-1}(z)$.
}
$\partial_zg(z)=\xi_\ast^{-1}(z)$.
Therefore the saddle point equations can be inverted to give 
\eq{
\tilde{p}(\xi):=
p(\xi) - p(\xi_0)= g(z_\ast(\xi))-\frac{z_\ast(\xi)}{\xi}\,,
\label{eq:pxi-sp}
}
with $z_\ast(\xi)$ given by the solution to 
$\partial_z g(z)\big\vert_{z=z_\ast(\xi)} = 1/\xi$, i.e.\
(via Eq.~\ref{eq:gz}) $\xi_{\rm typ}^{-1} -2bz_\ast(\xi)=\xi^{-1}$ whence
\eq{z_\ast(\xi) = (\xi_{\rm typ}^{-1}-\xi^{-1})/2b\,.\label{eq:zstar}}

At this point it is useful to note that $g(z)$ is a concave function with a positive gradient for 
$z\in[0,1]$. In particular, $\xi_{\rm max}^{-1}\leq g'(z)\leq \xi_{\rm typ}^{-1}$ where $\xi_{\rm max}^{-1} = \xi_{\rm typ}^{-1}-2b$. Physically, this means that while the distribution
$P_\xi$ in principle has support for all $\xi>0$, only the domain $\xi_{\rm typ}\leq\xi\leq\xi_{\rm max}$ is physically relevant for determining $c_{\rm avg}^{(q)}(r,L)$ via Eq.~\eqref{eq:cavg-Pxi}; which in effect is thus blind to the behaviour of $P_\xi$ outside this $\xi$-range.
For $P_{\xi} \sim\exp[-L\tilde{p}(\xi)]$, Eqs.~\eqref{eq:pxi-sp},\eqref{eq:gz} give
$\tilde{p}(\xi)=bz_{*}^{2}(\xi)$ and hence 
\eq{
P_\xi(\xi,L) \sim \exp\left[-\frac{L}{4b}\left(\xi_{\rm typ}^{-1}-\xi^{-1}\right)^2\right]\,,
\label{eq:P-xi}
}
which of course holds over the interval $\xi\in[\xi_{\rm typ},\xi_{\rm max}]$.

A few physical remarks are in order. The form of  Eq.~\eqref{eq:P-xi} means that the distribution itself is well behaved and not fat-tailed. This implies that rare pairs of Fock-space nodes, which dominate the average correlations, arise from the interplay of such a distribution with the exponentially large number of such pairs of nodes.
In addition, it is important to note that these pairs arise entirely due to 
interactions, namely to the first term in the Hamiltonian in Eq.~\eqref{eq:hamiltonian}.
This is evinced by the fact that in the non-interacting limit, $J_i=0$, both the average and typical correlation functions can be shown to decay purely exponentially with $r$ (of course with slightly different correlation lengths~\cite{roy2021fockspace}).
Finally, a more careful treatment of the normalisation factors in the steepest-descent analysis shows that for $L\to\infty$, the distribution reduces to $P_\xi=\delta(\xi-\xi_{\rm typ})$, consistent with the variance vanishing with $L$. This further  underscores the key role played by the high-dimensional nature of the Fock-space graph, in the average correlation functions being dominated by rare pairs of Fock-space nodes.  The form of the distribution in Eq.~\ref{eq:P-xi} 
shows that ${\rm var}[\xi]$ decreases $\propto L^{-1}$. However, for a given eigenstate, the average correlation function $c_{\rm avg}^{(q)}(r,L)$ samples the well-behaved `tails' of the distribution exponentially many times, allowing for the emergence of the rare pairs.

\begin{figure}
    \centering
    \includegraphics[width=\linewidth]{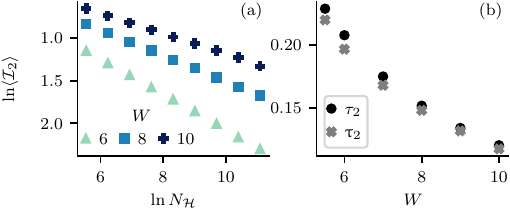}
    \caption{(a) Scaling of the IPR $\langle \mathcal{I}_{2}\rangle$ with the Fock-space dimension for different disorder strengths in the MBL phase. (b) 
		Multifractal exponent $\tau_2$, extracted from linear fits to the data in (a), compared with the analytical prediction [Eq.~\eqref{eq:tau2-corrs}] derived from the average correlation functions, denoted here by $\uptau_2$.
		}
    \label{fig:ipr2}
\end{figure}

\paragraph*{Correlation lengths and IPR:} 
Since $C_{\rm avg}^{(q)}({r=0})$ is directly connected to the generalised IPRs [see Eq.~\eqref{eq:Cavg-IPR}], it is instructive to determine how the correlation lengths control the latter. 
Exploiting the sum rule $\nh\sum_{r=0}^L N_r C_{\rm avg}^{(1)}(r,L)=1$, 
Eqs.~\eqref{eq:C-rescaled}-\eqref{eq:c-avg-typ-scaling} give
\eq{
\sum_{r=0}^L N_r^{\pd} e^{-Lg(r/L)} = \braket{{\cal I}_2}^{-1}\sim e^{L\tau_2^{\pd} \ln 2}\,.
\label{eq:gx-IPR}
}
Taking the density of nodes $N_{r}=\binom{L}{r} \simeq e^{LH(x)}$ with $H(x)$ the binary entropy
function, Eq.~\eqref{eq:gx-IPR} can be evaluated via steepest descents. 
Since $g(x)$ is quadratic in $x$, the saddle-point equation lacks a closed-form solution. 
Instead, the result is parametrically expressed as~\cite{supp} 
\eq{
\tau_2^{\pd} \ln 2 = -\ln(1-x_\ast^{\pd}) - b x_\ast^2\,,
}
where the saddle point $x_\ast$ satisfies
\eq{
\xi_{\rm typ}^{-1} = 2bx_\ast^{\pd} - \ln\left(\frac{x_\ast}{1-x_\ast}\right)\,.
}
Numerically, we find~\cite{supp} that $x_\ast\ll 1$ deep in the MBL phase,  which permits an analytical approximation to the saddle point. With $x_\ast\ll 1$, one can show that~\cite{supp}
\eq{
\begin{split}
&\tau_2^{\pd}\ln 2 \simeq \ln A - \ln(1-2b\zeta) - \frac{b\zeta^2}{A^2}\,,\\ 
&{\rm with}\quad A = 1-(2b-1)\zeta \quad{\rm and}\quad \zeta = e^{-\xi_{\rm typ}^{-1}}\,.
\end{split}
\label{eq:tau2-corrs}
}
This establishes an explicit, albeit approximate, relationship between the multifractal exponent $\tau_2$, and $\xi_{\rm typ}$ and $b$ which govern the Fock-space correlations. 
As $W$ increases, $\xi_{\rm typ}(W)$ naturally decreases, leading to a suppression of $\tau_2$ in accordance with Eq.~\eqref{eq:tau2-corrs}.
In Fig.~\ref{fig:ipr2}, we compare the multifractal exponent extracted directly from power-law fits of $\langle{\cal I}_2\rangle$ with $\nh$, to the analytical prediction from the right-hand side of Eq.~\eqref{eq:tau2-corrs}. The results are in excellent agreement, 
validating our correlation-length distribution theory.

\paragraph*{Singularity spectrum and anomalous dimension:} 
Building on the explicit relationship between the correlation functions and 
$\langle{\cal I}_2\rangle$, it is natural to extend this connection to the generalised IPRs, 
$\langle{\cal I}_q\rangle$, and the singularity spectrum, $f(\alpha)$, of the MBL eigenstates, defined via~\cite{evers2008anderson}
\eq{
\braket{{\cal I}_q} = 
\sum_{I}\braket{\abs{\psi_I^{\pd}}^{2q}} = \int d\alpha\, \nh^{-\alpha q + f(\alpha)} \sim \nh^{-\tau_q}\,.
\label{eq:Iq}
}
The exponent $\tau_q$ is related to the anomalous dimension $\Delta_q$
by $\Delta_q=\tau_q-q+1$~\cite{evers2008anderson}. For our numerical analysis, we define the distribution of the logarithmic eigenstate amplitudes as 
\eq{
P_\alpha(\alpha, L) = \frac{1}{\nh}\sum_I \left\langle \delta\left(\alpha + \frac{\ln\abs{\psi_I}^2}{L\ln 2}\right) \right\rangle\,,
\label{eq:P-alpha-def}
}
which is directly related to the singularity spectrum via $P_\alpha(\alpha,L) = \exp[(f(\alpha)-1)L\ln 2]$.

\begin{figure}
\includegraphics[width=\linewidth]{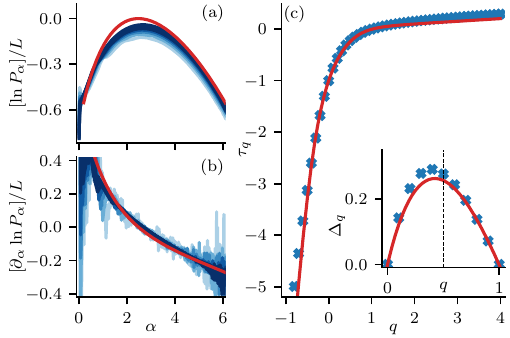}
\caption{(a) The logarithm of the distribution $P_\alpha$, rescaled by $L$, and (b) its derivative with respect to $\alpha$ for $L=8,\cdots,16$ (lighter to darker colours). Red lines denote the theoretical predictions from Eqs.~\eqref{eq:yogibear},\eqref{eq:pi-alpha}. (c) The multifractal exponent $\tau_q$ as a function of $q$, where points represent numerical data obtained from the scaling of $\langle{\cal I}_q\rangle$ with $\nh$, and the red line denotes the theoretical curve from Eq.~\eqref{eq:tauq-ana}. The inset shows the anomalous dimension
$\Delta_{q}$, clearly highlighting the asymmetry $\Delta_q \neq \Delta_{1-q}$ (the vertical dashed line is $q=1/2$). Data are for $W=8$ and $c= 0.69$. This same constant $c$ yields quantitatively accurate results for other values of $W$~\cite{supp}.}
\label{fig:Palpha}
\end{figure}

To connect the $P_\xi$ distribution to the singularity spectrum, we model the eigenstate amplitudes as 
\eq{
\abs{\psi_I^{\pd}}^2 = \mathcal{N}^{-1} e^{-r_{II_{0}}/\tilde{\xi}_I}
\,,
\label{eq:psi-model}
}
where the length scale $\tilde{\xi}$ is proportional to the correlation length, $\tilde{\xi} = c\,\xi$, with a constant $c$ independent of $L$. Here, $I_{0}\equiv I_{\rm max}$ denotes the localisation centre, identified as the Fock-space node with the maximum amplitude. 
Because the number of nodes at any given distance from $I_{0}$ grows exponentially with $L$, we 
assume that the amplitudes of a single eigenstate self-average across the Fock-space.
Consequently, the normalisation becomes
\eq{
{\cal N} \simeq \sum_I \int d\xi_I^{\pd}\,e^{-L\tilde{p}(\xi_I^{\pd})}\,e^{-\frac{r_{II_{0}}^{\pd}}{c\xi_I}} 
~\sim ~ e^{-L\eta_\ast^{\pd}}\,.
\label{eq:wf-norma}
}
Employing this description in Eq.~\eqref{eq:P-alpha-def}, one can show~\cite{supp} that $P_\alpha$ assumes the large-deviation form
\eq{
P_\alpha(\alpha,L) \sim \exp[L\pi(\alpha)]\,,
\label{eq:yogibear}
}
where 
\eq{
\pi(\alpha) = \max_{0\leq x\leq 1} [H(x)-\tilde{p}(\xi_\ast(x,\alpha))-\ln 2]\,,
\label{eq:pi-alpha}
}
with $\xi_\ast(x,\alpha) = x/[c(\alpha \ln 2 + \eta_\ast)]$. Importantly, this implies that the distribution of correlation lengths—parametrised by $\xi_{\rm typ}$ and $b$—predicts a concrete analytical form for $P_\alpha$. In Fig.~\ref{fig:Palpha}(a), we compare this prediction with exact numerical data. The finite-size drift visible in the raw data originates from the $L$-dependent normalisation factor in $P_\alpha$. To circumvent this, we compute the derivative 
$[\partial_\alpha \ln P_\alpha(\alpha,L)]/L$, which exhibits good convergence with $L$. As shown in Fig.~\ref{fig:Palpha}(b), the theoretical prediction from Eqs.~\eqref{eq:pi-alpha},\eqref{eq:yogibear} is in excellent agreement with this  data.

An analogous calculation using Eqs.~\eqref{eq:psi-model} and \eqref{eq:wf-norma} in Eq.~\eqref{eq:Iq} also yields a theoretical prediction for the $\tau_q$ exponent~\cite{supp}:
\eq{
\tau_q^{\pd} \ln 2 = \min_{0\leq x\leq 1}\big[g(xq/c) - H(x) - q\eta_\ast^{\pd}\big]\,.
\label{eq:tauq-ana}
}
As seen in Fig.~\ref{fig:Palpha}(c), this too describes very well the $\tau_q$ extracted directly from the numerical data.
An important consequence of this framework is that it theoretically captures the symmetry violation of quantum multifractality, $\Delta_q \neq \Delta_{1-q}$, as visually evident in the inset to Fig.~\ref{fig:Palpha}(c).
From Eq.~\eqref{eq:tauq-ana}, one can straightforwardly show that 
\eq{
\partial_q^{\pd}\Delta_q^{\pd}\big\vert_{q=\frac{1}{2}} = \frac{1}{\ln 2}\left[\frac{x_\ast(\frac{1}{2})}{c}g'\left(\frac{x_\ast(\frac{1}{2})}{2c}\right)-\eta_\ast\right]-1\,,
\label{eq:Delta-deriv}
}
where $x_\ast(q)$ minimises the right-hand side of Eq.~\eqref{eq:tauq-ana}. 
Because this derivative is generally non-zero, the symmetry $\Delta_q = \Delta_{1-q}$
manifestly does not hold. Physically, this  persists even if the correlation length distribution $P_\xi$ is tightly peaked as a Dirac $\delta$-function. This indicates that the violation stems entirely from the high-dimensional nature of the Fock-space graph~\cite{bilen2021symmetry}.


\paragraph*{Discussion:}
In summary, we developed a quantitative theory for the statistics of Fock-space correlation lengths in the MBL phase, as a route towards understanding the strong inhomogeneities of MBL eigenstates on the Fock-space.  While the distribution of these length scales is perfectly well-behaved, its interplay with the high-dimensional nature of the hypercubic Fock-space graph leads to qualitatively different behaviours of the average and typical Fock-space correlations. 
The demonstration of an $L$-independent $\xi_{\rm typ}$ is a clear signature of the stability of the MBL phase -- and how it behaves as the MBL transition is approached is certainly a question of immediate interest.

While in this work we focussed entirely on the eigenstates, it will be interesting to ask if similar physics manifests itself in the dynamics of time-evolving states on the Fock-space graph. In particular, it remains to be seen whether the dynamics exhibit
something akin to dynamical heterogeneity analogous to that of glass formers, but strictly on the Fock space. In addition, the average and typical Fock-space correlations can be related to correlations of real-space observables, even if non-local~\cite{supp}, as a step towards establishing a concrete connection between the Fock-space inhomogeneities and long-ranged resonances. Finally, it would be exciting and timely to understand these inhomogeneities from the viewpoint of wavefunction networks~\cite{mendessantos2024wave,andreoni2025network}, with the hope of probing them experimentally in the not-too-distant future~\cite{arute2019quantum,lunkin2026multifractal}.

\begin{acknowledgments}
{\it Acknowledgements:} We thank A. Scardicchio for helpful discussions. This work was supported by the Department of Atomic Energy, Government of India, under project nos. RTI4019 and RTI4013. S.R. acknowledges support from SERB-DST (India) under Grant No.
SRG/2023/000858, from ANRF (India) under Grant No.
ANRF/ARG/2025/004045/PS, and from a Max Planck
Partner Group grant between ICTS-TIFR, Bengaluru
and MPIPKS, Dresden.
\end{acknowledgments}

\bibliography{references}

\clearpage

\onecolumngrid

\begin{center}
{\large{\bf Supplementary Material:\mytitle}}\\
\medskip

Bibek Saha\textsuperscript{1}, David E. Logan\textsuperscript{2}, and Sthitadhi Roy\textsuperscript{1}\\
\smallskip

{\small{\it \textsuperscript{1}International Centre for Theoretical Sciences, Tata Institute of Fundamental Research, Bengaluru 560089, India\\
\textsuperscript{2}University of Oxford, Physical and Theoretical Chemistry, South Parks Road, Oxford OX1 3QZ, UK}}

\end{center}


\section{IPR exponent and Fock-space correlation functions}

Here we provide details of the calculations relating the  IPR exponent $\tau_2$ to 
$\xi_{\rm typ}$ and $b$ which control the correlation functions. To this end note that the sum in Eq.~\eqref{eq:gx-IPR} can be recast as an integral
\eq{
\int_0^1 dx\, e^{L[H(x)-g(x)]}\sim e^{L\tau_2^{\pd}\ln2}\,,
\label{eq:gx-IPR-int}
}
using Stirling's leading approximation $N_{r}=\binom{L}{r}=\binom{L}{Lx} \simeq e^{LH(x)}$, with
\eq{
H(x) = -x\ln x -(1-x)\ln(1-x)\,
\label{eq:Hx}
}
the binary entropy function. Using a saddle-point approximation on Eq.~\eqref{eq:gx-IPR-int} gives
\eq{
\tau_2^{\pd}\ln 2 = \max_{x}[H(x)-g(x)] = H(x_\ast^{\pd})- g(x_\ast^{\pd})\,,
\label{eq:tau2-basic}
}
with the saddle point given by $[H'(x)-g'(x)]|_{x=x_\ast} = 0$. The form of $H(x)$ in Eq.~\eqref{eq:Hx} and $g(x)$ in Eq.~\eqref{eq:gz} leads to a transcendental equation for the saddle point, 
\eq{
\xi_{\rm typ}^{-1} = 2bx_*^{\pd} - \ln\left(\frac{x_*}{1-x_*}\right)\,. \label{eq:saddle-point-trans}
}
Using the saddle point back in Eq.~\eqref{eq:tau2-basic} then yields the parametric result
\eq{
\tau_2^{\pd} \ln 2 = -\ln(1-x_*^{\pd}) - bx_*^2\,.\label{eq:tau2-sp}
}

\begin{figure}
    \centering
\includegraphics[width=0.5\linewidth]{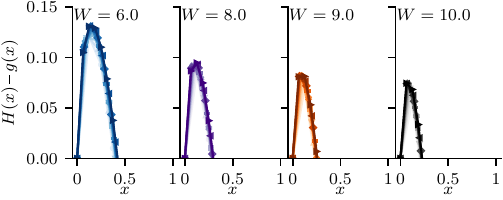}
    \caption{The function $H(x)-g(x)$ which appears in the saddle-point approximation of $\tau_2$ [see Eq.~\eqref{eq:tau2-basic}]. The location of the maximum of this function, $x_\ast$, is the saddle point which governs $\tau_2$. Data is given for $L = 8, 9, \cdots 16$ (darker intensities denote larger $L$), and clearly show that $x_\ast\ll 1$ for all $W$  
considered in the MBL phase.
 }
    \label{fig:sp-curve}
\end{figure}

At this stage we  assume that for disorder strengths $W$ deep in the MBL phase, 
the saddle point $x_\ast \ll 1$;  which is indeed corroborated numerically in 
Fig.~\ref{fig:sp-curve}. Within this approximation,  we have from Eq.~\eqref{eq:saddle-point-trans}
\eq{
\xi_{\rm typ}^{-1} \simeq (2b-1)x_*^{\pd} - \ln x_\ast^{\pd}\,,
}
which in turn implies 
\eq{
x_\ast \simeq e^{-\xi_{\rm typ}^{-1}}e^{(2b-1)x_\ast^{\pd}}\,.
}
Again using $x_\ast \ll 1$ to linearise the exponential above  leads to a solution for 
the saddle point,
\eq{
x_\ast \simeq \frac{e^{-\xi_{\rm typ}^{-1}}}{1-e^{-\xi_{\rm typ}^{-1}}(2b-1)}\,.
}
Inserting the above solution into Eq.~\ref{eq:tau2-sp} finally leads to
\eq{
\tau_2 \ln 2 \simeq  \ln\left[1 - (2b-1)e^{-\xi_{\rm typ}^{-1}}\right] - \ln\left(1 - 2b e^{-\xi_{\rm typ}^{-1}}\right) - \frac{b e^{-2\xi_{\rm typ}^{-1}}}{\left[1 - (2b-1)e^{-\xi_{\rm typ}^{-1}}\right]^2}\,,
}
which is the result in Eq.~\eqref{eq:tau2-corrs}.

\section{Details of the calculation of \boldmath{$P_\alpha(\alpha,L)$} and \boldmath{$\tau$}$_q$}

Here we give details of the calculation of $P_\alpha(\alpha,L)$. As a starting point, we express the eigenstate amplitudes for a given eigenfunction as 
\eq{
\abs{\psi_{I}^{\pd}}^2 = {\cal N}^{-1}~e^{-r_{I_0I}^{\pd}/\tilde{\xi}_I^{\pd}}\,,
\label{eq:psi-K-model}
}
where $I_0$ is the localisation centre of the eigenstate on the Fock-space graph, identified by the maximum of $\abs{\psi_I}^2$ over all the Fock-space nodes. 
Here, while the lengthscale $\tilde{\xi}$ need not necessarily be the same as the correlation length $\xi$~\cite{roy2021fockspace}, we expect it to be proportional to $\xi$,
and as mentioned in the main text thus take
\eq{
\tilde{\xi} = c\,\xi\,
}
where $c$ is a constant which does not depend on system size.

The normalisation factor is given by ${\cal N} = \sum_I e^{-r_{I_0I}/\tilde{\xi}_I}$. 
Since the number of Fock-space nodes at distance $r=xL$ from $I_0$ is 
in general exponentially large in $L$,
the summand in the normalisation can be replaced by an integral over the probability distribution 
$P_\xi(\xi,L)$.  In effect, this assumes the amplitudes self-average on the Fock space. This gives
\eq{
{\cal N} \simeq\sum_{I}\int d\xi~ P_{\xi}^{\pd}(\xi,L)
e^{-r_{I_{0}I}^{\pd}/c\xi}
~=~ \sum_{I}e^{-Lg(x_{I_0I}^{\pd}/c)}
\,,
}
where we simply use the average correlation function given from  
Eqs.~\eqref{eq:cavg-Pxi},\eqref{eq:c-avg-typ-scaling}.
The sum over $I$ can again be done via a saddle-point treatment, which yields
\eq{
{\cal N}\simeq \int_0^1 dx\,e^{-L[-H(x)+g(x/c)]}~\sim~ e^{-L\eta_\ast}\,
\label{eq:norma-sm}
}
where $\eta_\ast = \min_x[\eta(x,c)]$ with $\eta(x,c)=-H(x)+g(x/c)$.
With this normalisation, the exponent $\alpha$ for a given Fock-space node is 
\eq{
\alpha_I^{\pd} = -\frac{\ln |\psi_I^{\pd}|^2}{\ln\nh} = \frac{x_{I_0I}^{\pd}}{c\xi_I^{\pd}\ln 2} - \frac{\eta_\ast^{\pd}}{\ln 2}\,.
}
Using this form in Eq.~\eqref{eq:P-alpha-def}, and again assuming 
self-averaging as above, we obtain
\eq{
P_\alpha^{\pd}(\alpha,L) \simeq
\frac{1}{\nh}\sum_I \int d\xi~P_\xi^{\pd}(\xi,L)~\delta\left(\alpha - \frac{x_{I_0I}^{\pd}}{c\xi\ln 2} + \frac{\eta_\ast^{\pd}}{\ln 2}\right).
}
With $P_{\xi}(\xi,L) \sim e^{-L\tilde{p}(\xi)}$ as per the main text, and again
using Stirling's approximation for $N_{r}$, this gives
\eq{
P_\alpha^{\pd}(\alpha,L) \simeq 
\int_0^1 dx ~ e^{L[H(x) - \tilde{p}(\xi_\ast^{\pd}(x,\alpha))-\ln 2]}
\label{eq:Palpha-sp}
}
where 
\eq{
\xi_\ast^{\pd}(x,\alpha) = \frac{x}{c[\alpha \ln 2 + \eta_\ast^{\pd}]}\,.
}
A saddle-point approximation to the integral in Eq.~\eqref{eq:Palpha-sp} then leads to
\eq{
P_\alpha^{\pd}(\alpha,L)\sim \exp[L\pi(\alpha)]\,;~~~~~~~\pi(\alpha) = \max_{0\leq x\leq 1} [H(x)-\tilde{p}(\xi_\ast^{\pd}(x,\alpha))-\ln 2]\,.
}
This shows explicitly the large-deviation form of the distribution $P_\alpha(\alpha,L)$.

\begin{figure}
\includegraphics[width=0.5\linewidth]{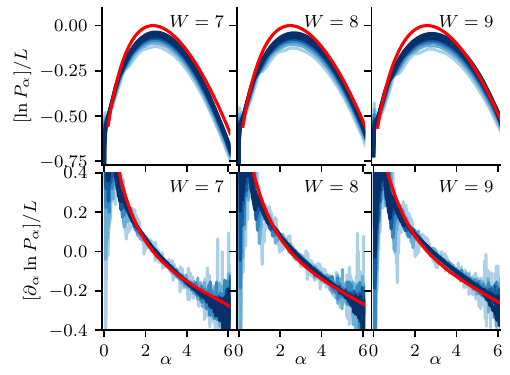}
\caption{The analogue of Fig.~\ref{fig:Palpha}(a)-(b) of the main text, but shown in addition for $W=7$ and 9. The theoretical results (red curves) are obtained with the same constant $c\simeq 0.69$ for all $W$ values, indicating the validity of the theory over a broad range of disorder strengths in the MBL phase.}
\label{fig:Palpha-SM}
\end{figure}

To compare the numerically determined $P_\alpha$ (obtained from exact diagonalisation) with the theoretical prediction, we use the form of the $\tilde{p}(\xi)$ in Eq.~\eqref{eq:P-xi}, with 
$\xi_{\rm typ}$ and $b$ obtained from the Fock-space correlation functions. 
The residual $L$-dependence in the numerical data is due to the $L$-dependence in the normalisation of $P_{\alpha}$, which is subleading for $\ln P_\alpha$.
Since the normalisation does not however depend on $\alpha$, we also study $[\partial_\alpha \ln P_\alpha(\alpha,L)]/L$, which is well converged with $L$.
The results are given in Fig.~\ref{fig:Palpha-SM}, which shows remarkable agreement between the numerical data and the theoretical prediction. 

The same description as in Eq.~\eqref{eq:psi-K-model}, with the normalisation in Eq.~\eqref{eq:norma-sm}, can be used to compute the generalised IPR. From the definition 
$\langle{\cal I}_q\rangle = \sum_{I}\langle |\psi_{I}|^{2q}\rangle \sim \nh^{-\tau_{q}}$ 
(Eq.~\eqref{eq:Iq}) one has
\eq{
\langle{\cal I}_q\rangle \simeq \sum_I \int d\xi~P_\xi^{\pd}(\xi,L)~ \frac{e^{-\frac{qr_{I_0I}^{\pd}}{c\xi}}}{e^{-Lq\eta_\ast^{\pd}}} ~=~ \sum_I e^{-L[g(qx_{I_0I}^{\pd}/c)-q\eta_\ast^{\pd}]}\,,
}
where again we use the average correlation function given from 
Eqs.~\eqref{eq:cavg-Pxi},\eqref{eq:c-avg-typ-scaling}.
As before, using $\binom{L}{r} \simeq e^{LH(r/L)}$, the sum over $I$ can be converted to an integral and analysed using steepest descents; giving
\eq{
\tau_q^{\pd} \ln 2 = \min_{0\leq x\leq 1}\big[g(xq/c) - H(x) - q\eta_\ast^{\pd}\big]\,,
}
which is the result in Eq.~\eqref{eq:tauq-ana}.


\section{Connection between Fock-space correlation functions and real-space correlation functions}
In this section, we discuss the concrete relations between the Fock-space correlation functions discussed in the main text, and speculate on how they may be measured. For a given 
$\sigma^z$-configuration $|I\rangle$, consider the projector onto it given by
\eq{
\Pi_{I} = \bigotimes_{i=1}^L \left(\frac{\mathbb{I}+ s_i^I \sigma^z_i}{2}\right)\,,
}
where $s_i^I =\langle I|\sigma^{z}_{i}|I\rangle =\pm 1$ is the spin configuration at real-space site $i$ in the state $\ket{I}$. Next, consider a subsystem ${\cal S}_r$ of $r$ spins, not necessarily contiguous, and define the spin-flip operator on the subsystem as 
\eq{
X_{{\cal S}_r} = \bigotimes_{i \in {\cal S}_r}\sigma^x_i\,.
}
It is then straightforward to show that the squared expectation value of the operator ${\cal O}(I,{\cal S}_r) \equiv X_{{\cal S}_r}\Pi_I$ is related to the Fock-space correlation as
\eq{
\braket{\psi|{\cal O}(I,{\cal S}_r)|\psi}^2 = |\psi_I|^2|\psi_{I_{{\cal S}_r}}|^2\,,
}
where $\ket{I_{{\cal S}_r}}$ is the state obtained by flipping the spins in ${\cal S}_r$ in state $\ket{I}$.
Finally, note that the square of the expectation value can be achieved by measuring the autocorrelation of the operator ${\cal O}(I,{\cal S}_r)$ at late times,
\eq{
\braket{\psi|{\cal O}(I,{\cal S}_r)|\psi}^2 = \lim_{t\to\infty}\braket{\psi|e^{iHt}{\cal O}(I,{\cal S}_r)e^{-iHt}{\cal O}(I,{\cal S}_r)|\psi}\,.
}
Averaging over both ${I}$ and the subsystems ${\cal S}_r$ of size $r$ is equivalent to tracking the average Fock-space correlation $C_{\rm avg}^{q=1}(r,L)$, whereas randomly selecting a ${\cal S}_r$ and averaging over $I$ can yield the typical correlation $C_{\rm typ}^{q=1}(r,L)$.
This shows that the Fock-space correlations can indeed be constructed concretely from real-space correlations, albeit of non-local operators.

From an experimental standpoint, extracting these quantities on programmable quantum processors—such as superconducting qubit platforms—is  feasible. Modern architectures natively employ multiplexed dispersive readout to perform simultaneous, high-fidelity projective measurements across all qubits in the computational ($\sigma^z$) basis~\cite{arute2019quantum}. Each experimental ``shot" collapses the system into a specific bitstring $\ket{I}$ with probability $p_I = |\psi_I|^2$. Consequently, the correlations, 
$|\psi_I|^2|\psi_{I_{{\cal S}_r}}|^2$ corresponds precisely to the joint probability of sampling two specific bitstrings separated by a Hamming distance $r$. By repeating the state preparation and measurement cycle, these Fock-space correlations can be extracted directly via classical post-processing of the sampled distribution.

\end{document}